\documentclass[aps,pra,amsfonts,amssymb,amsmath,eqsecnum,twocolumn,showpacs]{revtex4}
\usepackage{graphicx}

\usepackage{color,graphicx,cancel}
\usepackage{psfrag}
\usepackage[utf8]{inputenc}
\usepackage[T2A]{fontenc}
\usepackage[russian,english]{babel}
\definecolor{darkblue}{rgb}{0,0,0.75}
\definecolor{darkred}{rgb}{0.7,0,0}
\definecolor{darkgreen}{rgb}{0.25,0.95,0}
\definecolor{darkbrawn}{rgb}{0.25,0,0.75}

\usepackage[unicode, colorlinks, citecolor=darkblue, linkcolor=darkred, urlcolor=blue]{hyperref}
\usepackage{euler}
\begin{document}

\title{Gyroscope as quantum angular speed meter}

\author{Andrey B. Matsko}
\affiliation{OEwaves Inc., 465 North Halstead Street, Suite 140, Pasadena, CA 91107}

\author{Sergey P. Vyatchanin}
\affiliation{Faculty of Physics, Moscow State University, Moscow 119991 Russia,\\
	    Quantum Technology Centre, Moscow State University, Moscow 119991 Russia}

\begin{abstract}
We found that the measurement sensitivity of an optical integrating gyroscope is fundamentally limited due to ponderomotive action of the light leading to the standard quantum limit of the rotation angle detection. The uncorrelated quantum fluctuations of power of clockwise and counterclockwise electromagnetic waves result in optical power-dependent uncertainty of the angular gyroscope position. We also show that, on the other hand, a quantum back action evading measurement of angular momentum of a gyroscope becomes feasible if proper measurement strategy is selected. The angle is perturbed in this case. This observation hints on fundamental inequivalency of integrating and rate gyroscopes.
\end{abstract}


\maketitle

There are two types of optical gyroscopes with respect to measurement observables -- rate and integrating ones. Rate gyroscopes, for instance, passive fiber optic gyroscopes, measure rotation speed $\Omega$. Integrating gyroscopes, for instance, laser gyroscopes, allow a direct observation of the rotation phase, $\phi$. Both devices are based on optical Sagnac effect and seem to be identical from the classical physics perspective. From quantum physics perspective, these measurements are not equivalent because angular momentum and rotation phase operators do not commute. In this paper we study fundamental quantum limitations of rotation measurement accuracy using a passive resonant optical resonant gyroscope as a model. We also study quantum limitations of the torque measurement using the system.

Passive resonant optical gyroscopes utilize the rotation-dependent change of the distance travelled by light in clockwise (cw) and counter clockwise (ccw) directions with respect to the rotation axis to measure the rotation rate and rotation angle. The non-reciprocity of the rotating system results in the removal of the frequency degeneracy between cw and ccw modes of the ring cavity. The frequencies of the modes can be measured in various ways, for instance by observation of the phase shift of the cw and ccw light interacting with them. The resonant nature of the device is important since the signal is proportional to the finesse of the cavity.

Usual quantum analysis of a gyroscope involves study of the optical quantum noise of the device \cite{dorschner80jqe,scully93pra}. The noise results in sensitivity limitation that can be lifted with increase of the optical pump power. It was noticed that sensitivity of a gyroscope can be also limited due to ponderomotive effect \cite{caves81prd}. It was envisioned that, in case the device has movable parts, like mirrors, the optical power fluctuations at the mirrors introduce additional noise term, called quantum back action \cite{braginsky80s}, that eventually limits the overall sensitivity of the gyroscope for any optical power, similarly to limitation of sensitivity of any interferometric measurement of this kind. Miniaturization of a gyroscope made out of a nonlinear material also results in the back action \cite{matsko18pla}. In this paper we study the fundamental limitations of the sensitivity of a macroscopic gyroscope and introduce uncertainty principle for the angular momentum and phase of a gyroscope. From this uncertainty principle we derive standard quantum limit for the rotation measurements and consider several practical examples of detection of initial angular velocity of a solid body as well as a torque acting on the body that lead to the standard quantum limit. A possibility of a back action evading measurement is discussed. We also introduce an analogy between the gyroscope and a device measuring speed of a solid state body (speed meter).

\section{Standard quantum limit of rotation angle}

Let us consider rotation of a rigid thin circular thread around axis $Z$. The motion can be described by angular momentum and rotation phase operators in analogy with momentum and coordinate of a point-like particle. The measurement of the rotation speed corresponds to the measurement of angular momentum $p_\phi=m r^2\Omega$, where $m$ is mass and $r$ is radius of the circle. The quantum operators corresponding to angular momentum and the phase, $\hat p_\phi=-i\hbar \partial/(\partial \hat \phi)$ and $\hat \phi$, do not commute,
\begin{equation} \label{commutator}
[\hat p_\phi, \hat \phi]=-i\hbar,
\end{equation}
which means that both the observables cannot be measured accurately at the same time and, hence, the rate and integrating gyroscopes are not equivalent. The Heisenberg's uncertainty principle leads to the requirement $\Delta_{p} \Delta_\phi \geq \hbar/2$, which also can be rewritten in form $\Delta_\Omega \Delta_\phi \geq \hbar/2 m r^2$. Here $\Delta_{p} $, $\Delta_\Omega$, and $\Delta \phi$ are uncertainties of the momentum, frequency, and phase. The expressions can be generalized for a 3D object, however we leave this exercise outside of the paper, for the sake of simplicity.

In addition to this inequivalency, the noncommutativity of the operators results in appearance of the effect of quantum back action in the rotation measurements. The back action leads to appearance of the Standard Quantum Limit (SQL) of measurements written as
\begin{eqnarray} \label{phisql}
\Delta \phi_{SQL} = \sqrt\frac{\hslash t_m}{mr^2}\\
\Delta \Omega_{SQL}= \sqrt\frac{\hslash }{mr^2 t_m} \label{omsql}
\end{eqnarray}
where we assume the simplest case of two simultaneous measurement of angle $\phi$ separated by time $ t_m $ \cite{braginsky68, braginskykhalili_book}. The SQL for the rotation angle has a complete analogy with corresponding to SQL of coordinate and momentum of a free mass.

The SQL value is small in practical devices, but can become measurable for a micro- or nano-device. For instance, $\Delta \phi_{SQL}=1.8 \times 10^{-5}$~degrees for $m=10^{-5}$~g, $r=10^{-2}$~cm, and $t_m=10^5$~s. The bias drift of a good gyroscope is two orders of magnitude larger for the same time of observation. On the other hand, assuming that the trend of technology development of smart nano-structures, we can see that there is a certain limit of miniaturization when the fundamental quantum limitations become important.

Equations (\ref{phisql}) and (\ref{omsql}) also show that the measurements of the rotation phase and frequency are fundamentally inequivalent. It also means that true integrating gyroscope and a rate gyroscope are also fundamentally different. The difference can be visible only on  quantum level, which is too low for the vast majority of practical applications.

\section{Standard quantum limit of rotation angle for an optical gyroscope}

\subsection{Model of the gyroscope}

Let us explain the origin of the SQL of the rotation angle using an example of a resonant optical gyroscope. An resonant optical gyroscope measures either angular speed or rotation angle, depending on the configuration. The measurements are based on Sagnac effect that results in frequency shift of a ring cavity mode as a function of the rotation frequency. In the case of clockwise (cw) rotation of the cavity the clockwise and counterclockwise (cw and ccw) frequencies of the cavity modes ($\omega_+$ and $\omega_-$, respectively) shift as
\begin{equation} \label{split}
\Delta \omega_\pm = \mp \frac{r\Omega}{cn_0},
\end{equation}
where $c$ is speed of light in the vacuum and $n_0$ is the refractive index of the material. A gyroscope can detect the Sagnac effect-mediated phase shift of the light passing through the rotating cavity, the frequency shift of light generated in the resonator filled with a lasing medium, or a fringe shift resulting from the interference of  cw and ccw light emitted by the gyroscope cavity. In the first two cases the gyroscope measures rotation speed. In the third case it measures the rotation angle.

To derive Eq.~(\ref{split}) we use standard formalism describing modes of a rotating optical cavity \cite{malykin14pu,matsko06oc}. At this point we neglect the mechanical degree of freedom and consider only optical part of the system. We assume that phase velocity of light in the motionless cavity is $V_{ph}=c/n_0$. For the rotating  cavity the optical path increases in the cw direction (denoted with subscript $+$) and decreases in the ccw direction (denoted with subscript $-$) in accordance with kinematic formulas
\begin{align}
 L_\pm &= 2\pi r \pm r\Omega\, t_\pm,\\
 V_\pm &= \frac{V_{ph} \pm r\Omega}{1\pm V_{ph}r\Omega/c^2},
\end{align}
where $t_\pm$ is the cavity round trip travel time, and $V_\pm$ is the phase velocity of the cw and ccw waves in the laboratory frame of reference. Using definition of $t_\pm$ we find expression for the optical path length, $ L_\pm$, for the cw and ccw waves in the laboratory reference frame using special relativity formalism:
\begin{align}
 t_\pm & = \frac{ L_\pm}{V_\pm},\quad \Rightarrow\\
   L_\pm &= 2\pi r \pm r\Omega \cdot \frac{L_\pm\left(1\pm V_{ph}r\Omega/c^2\right)}{V_{ph} \pm r\Omega},\\
   \label{Lpm}
   L_\pm & =2\pi r\cdot \frac{V_{ph}\pm r\Omega}{V_{ph}\left(1- \left[ r\Omega/c \right]^2\right)}.
\end{align}
We can see from this expression that the optical length increases in the cw direction and decreases in the ccw direction when the resonator rotates in the cw direction.

Let us consider a motionless cavity made out of material with dielectric permittivity $\epsilon$ and magnetic permeability $\mu=1$. The refractive index of the cavity material is related to the dielectric permittivity $n_0 =\sqrt \epsilon$. Our goal is to find equations describing generalized canonical amplitudes $q_\pm(t)$ for cw and ccw modes of cavity.

We assume that the radius of cavity and effective cross section of the modes are large enough: $r\gg S^{1/2} \gg \lambda$ ($\lambda$ is a mean optical wavelength in the vacuum). In this case  electric ($E_\pm$) and magnetic ($H_\pm$) field amplitudes of the modes can be expressed as follows, see Sec. 83 in \cite{LL8},
\begin{subequations}
   \label{motionless}
 \begin{align}
   E_\pm(t) &= \mp \sqrt{\frac{2\pi  }{n S_\pm L_\pm}}
    \,f_\pm e^{\pm ik_\pm r\phi} \, \partial_t q_\pm(t),\\
   H_\pm(t) &= \sqrt{\frac{2\pi\, n }{ S_\pm L_\pm}}
   f_\pm e^{\pm ik_\pm r\phi} \, \omega_\pm q_\pm(t),\\
   f_\pm &=f_\pm(\vec r_\bot), \quad S_\pm=\int |f_\pm|^2\vec dr_\bot,\\
   k_\pm &= \frac{\omega_\pm}{V_{ph}}=\frac{n\omega_\pm}{c}, \quad V_{ph}=\frac{c}{n}
\end{align}
\end{subequations}
where $\phi$ is azimuthal angle, "$r\phi$" is the coordinate along the rim of ring cavity.

The translation condition $ E_\pm(t)= E_\pm(t+t_\pm)$ defines normal frequencies of the ring cavity. For the case of the cavity at rest ($\Omega=0$)
\begin{align}
  k_\pm &\cdot L_\pm = 2\pi \ell, \quad L_\pm = 2\pi r, \\
  \label{omega0}
 \Rightarrow& \quad \omega_\pm= \ell\, \frac{V_{ph}}{r}\equiv \omega_0,
\end{align}
where $\ell$ is integer. Normal frequencies of the cw and ccw modes are degenerate $\omega_+ =\omega_-$ and are equal to  $\omega_0$.

Kinetic, $T_\pm$, and potential, $U_\pm$, energies for the optical fields and associated Lagrangians $\mathcal L_\pm$ can be expressed in terms of the canonical amplitudes as follows
\begin{align}
   T_\pm &=\int \frac{L_\pm\epsilon\langle |E_\pm|\rangle^2 }{8\pi}\, dr_\bot = \frac{\partial_t q_\pm^2}{2},\\
   U_\pm  &=  \int \frac{L_\pm\langle |H_\pm|\rangle^2)}{8\pi}\, dr_\bot = \frac{\omega_\pm ^2q_\pm^2}{2},\\
   \label{L0}
   \mathcal L_\pm &= T_\pm -U_\pm = \frac{\partial_t q_\pm^2}{2} - \frac{\omega_\pm ^2q_\pm^2}{2}.
\end{align}
We can find generalized momentums, $p_\pm$, for the light confined in the modes and write down corresponding Hamiltonians, $H_\pm$, for the optical fields propagating in he cw and ccw directions
\begin{align}
 p_\pm & = \frac{\partial \mathcal L_\pm}{\partial t} = \partial_t q_\pm,\\
 H_\pm &= p_\pm \cdot \partial_t q_\pm -\mathcal L_\pm = \frac{p_\pm^2}{2} + \frac{\omega_\pm ^2q_\pm^2}{2}.
\end{align}

For gyroscope rotating with frequency $\dot \phi=\Omega$ the formulas \eqref{motionless} are valid. Using the translation conditions we obtain
\begin{align}
  k_\pm &\cdot L_\pm = 2\pi \ell, \quad \eqref{Lpm} \quad  \Rightarrow\\
  \omega_\pm &= 2\pi \ell\, \frac{V_{\pm}}{L_\pm}.
\end{align}
where $\omega_0$ is defined in \eqref{omega0}. Linearizing the expression with respect of  $(r\dot \phi/V_{ph})$ we derive
\begin{align}
  \omega_\pm \simeq \omega_0 \left ( 1 \mp \frac{r \dot \phi}{n_0 c} \right ),
\end{align}
that results in Eq.~(\ref{split}).

\subsection{Hamilton formalism for the opto-mechanical system}

We have characterized the field amplitudes of the optical modes of the ring cavity used in an optical gyroscope in a classical canonical way. To describe the measurement of the rotation using the gyroscope in quantum picture we need to describe the interaction of light and mechanical degree of freedom in a classical canonical way and then quantize it. This can be done in two ways. We can either use Eq.~(\ref{split}) and derive the complete Hamiltonian of the opto-mechanical system through Lagrangian formalism, or we can utilize Lagrangian of the electromagnetic modes given by Eq.~(\ref{L0}) directly, along with the Lagrangian of the mechanical system. The first method is similar to the standard opto-mechanical approach stating that the mechanical motion modifies the frequency of light and does not change the photon number. The second method is more general as it does not require photon number conservation in the modes but rather show that the number is conserved. Both methods result in the same final expression of the opto-mechanical Hamiltonian of a rotating system, and we present both derivations in what follows.

\subsubsection{Derivation of Hamiltonian using Eq.~(\ref{split})}

Let us consider a thin planar lossless fiber loop cavity of mass $m$ that can rotate. The loop confines monochromatic light in one of its cw/ccw mode pair. The kinetic and potential energy of the loop is
\begin{align}
T &=\frac{1}{2}mr^2 \dot \phi^2, \\
V &=E_+ \left [ 1- \frac{r \dot \phi}{cn_0}\right ]+
E_- \left [ 1+ \frac{r\dot \phi}{cn_0}\right ],
\end{align}
where we expressed optical energy stored in cw and ccw modes as $E_\pm=T_\pm+U_\pm$ when the cavity is at rest. The canonical angular momentum of the system is
\begin{equation}
p_\phi=\frac{\partial {\cal L}}{ \partial \dot \phi}= mr^2\dot \phi+ \frac{r}{cn_0} (E_+-E_-),
\end{equation}
where ${\cal L} = T-V$ is the Lagrangian. The Hamiltonian of the opto-mechanical system is defined as
\begin{align}
\label{H1}
H &=p_\phi \dot \phi - {\cal L}= \\ \nonumber
 &= \frac{1}{2I} \left (  p_\phi-\frac{r}{cn_0}(E_+-E_-) \right)^2+E_++E_-.
\end{align}
where $I=mr^2$ is the moment of inertia.

\subsubsection{Derivation of Hamiltonian using Eq.~(\ref{L0}) }

Usage of Eq.~(\ref{L0}) allows for a more generalized derivation of the Hamiltonian. We add a Lagrangian for a mechanical degree of freedom to the Lagrangian of light confined in the resonator modes to find a Lagrangian of the rotating opto-mechanical system
\begin{align}
 \mathcal L &= \mathcal L_m+ \mathcal L_+ +\mathcal L_- + \mathcal L_\text{int},\\
 \mathcal L_m &= \frac{I\,\dot \phi^2}{2},\quad
 \mathcal L_\pm = \frac{\partial_t q_\pm^2}{2} - \frac{\omega_0 ^2 q_\pm^2}{2},\\
 \mathcal L_\text{int} &= \omega_0^2 \frac{ r\dot \phi}{n_0c}\left(q_+^2-q_-^2\right)
\end{align}

Canonical angular momentum $p_\phi$ , as well as canonical optical momenta $\ p_+$  and $p_-$  of the system can be found from
\begin{align}
 p_\phi  &\equiv \frac{\partial \mathcal L}{\partial \dot \phi}
      = I\, \dot\phi+  \omega_0^2 \frac{ r}{n_0 c}\left(q_+^2-q_-^2\right),
      \\
    p_\pm & \equiv \frac{\partial \mathcal L}{\partial \dot q_\pm}  =\dot q_\pm, \\
   \dot\phi &= \frac{1}{I}\left(p_\phi  - \omega_0^2 \frac{ r}{n_0 c}\left(q_+^2-q_-^2\right)\right),
\end{align}
Hamiltonian of system is defined as
\begin{subequations}
\label{H2}
\begin{align}
 H &= p_\phi  \dot \phi + \dot q_+ p_+ + \dot q_-p_- - \mathcal L =\\
   &= H_+ + H_- + H_m   ,\\
   H_\pm & = \frac{ p_\pm^2}{2} + \frac{\omega_0 ^2 q_\pm^2}{2},\\
   H_m &=\frac{1}{2I}\left(p_\phi  - \omega_0^2 \frac{ r}{n_0 c}\left(q_+^2-q_-^2\right)\right)^2
\end{align}
\end{subequations}
\subsubsection{Equations of mothion}

Now using \eqref{H1} or \eqref{H2} one can write down equations of motion:
\begin{subequations}
 \begin{align}
  \partial_t q_\pm &= p_\pm,\\
  \partial_t p_\pm &= -\omega^2_0q_\pm\pm\\
  &\pm \frac{1}{I}\left(p_\phi  - \omega_0^2 \frac{ r}{n_0 c}\left(q_+^2-q_-^2\right)\right)
      2 \omega_0^2\cdot \frac{ R}{n c}\, q_\pm,\nonumber\\
  \partial_t p_\phi &=0,\\
  \partial_t\phi &=  \frac{1}{I}\left(p_\phi  - \omega_0^2\frac{ r}{n_0 c}\left(q_+^2-q_-^2\right)\right)
 \end{align}
\end{subequations}
These equations can be reduced to
\begin{subequations}
 \label{EqsH}
\begin{align}
  \partial_t^2 q_+  +\omega^2_0\left(1-
      2  \frac{ r}{n_0 c} \partial_t\phi \right)q_+ &=0,\\
   \partial_t^2 q_-  +\omega^2_0\left(1+
      2  \frac{ r}{n_0 c} \partial_t\phi \right)q_- &=0,\\
      I\partial_t^2 \phi + \omega_0^2 \frac{ r}{n_0 c}\partial_t \left(q_+^2-q_-^2\right)&=0
\end{align}
\end{subequations}

\subsection{Quantization}

At this point we are ready to quantize the rotating opto-mechanical system. Introducing annihilation and creation  operators for the optical modes ($\hat a_\pm^\dag$ and $\hat a_\pm$) we obtain
\begin{align}
 \hat q_\pm &= \sqrt\frac{\hslash }{2\omega_0}\left(\hat a_\pm + \hat a_\pm^\dag\right),\quad
      \hat p_\pm =\sqrt\frac{\hslash \omega_0}{2}\left(\frac{\hat a_\pm + \hat a_\pm^\dag}{i}\right),\nonumber\\
      \hat p_\phi & = mr^2\dot \phi+ \frac{\hbar \omega_0 r}{cn_0} \big(\hat a_+^\dag \hat a_+ -\hat a_-^\dag \hat a_-\big), \\ \hat H &= \hat H_m+ \hat H_+ +\hat H_-, \;\;
 \hat H_\pm = \hslash\omega_0\left(\hat a_\pm^\dag \hat a_\pm + \frac 1 2\right), \nonumber \\
 \hat H_m & =\frac{1}{2I}\left(\hat p_\phi  - \hslash\omega_0 \frac{ r}{n_0 c}
    \left(\hat a_+^\dag \hat a_+ -\hat a_-^\dag \hat a_-\right)\right)^2. \label{ham}
\end{align}
Here we used rotation wave approximation and dropped fast oscillating terms $\sim \hat a_\pm^2,\ (\hat a_\pm^\dag)^2$. To complete the picture we  need to take into account the commutator given by Eq.~(\ref{commutator}).

Equation~(\ref{ham}) immediately shows that the canonical angular momentum $\hat p_\phi$ is not perturbed by the interaction since $[\hat H, \hat p_\phi] = 0$. Therefore, $\hat p_\phi$ is conserved in the measurement. Same is related to the photon numbers. On the other hand, the rotation phase as well as optical phase is perturbed due to quantum back action.

To find expression for the quantum back action we write the Hamiltonian equations
\begin{subequations}
 \label{init}
 \begin{align}
  \dot {\hat p}_\phi & =\partial_\phi H = 0, \; \;\; \partial_t (\hat a_\pm^\dag \hat a_\pm )= 0, \\
  \label{dotphi}
  \dot {\hat \phi} & = \frac{\hat p_\phi}{I}-\frac{\hbar \omega_0 r}{I cn_0} (\hat a_+^\dag \hat a_+ -\hat a_-^\dag \hat a_-), \\
  \label{apm2}
  \dot {\hat a}_\pm & =
      -i\omega_0\left ( 1 \mp \frac{r \dot {\hat \phi}}{cn_0}  \right )\hat a_\pm,
\end{align}
\end{subequations}
that can be solved exactly, taking into account that $\hat a_\pm^\dag \hat a_\pm$ and $\hat p_\phi$ are conserved,
\begin{align}
\label{hatphi}
  \hat \phi &= \hat \phi_0+ \left (\frac{\hat p_\phi}{I}-\frac{\hbar \omega_0 r}{I cn_0}
    \big(\hat a_+^\dag \hat a_+ -\hat a_-^\dag \hat a_-\big) \right )t, \\
 {\hat a}_\pm &= e^{-i\omega_0 t}\times \\
  \times &\exp \left ( \pm i \frac{r }{cn_0}  \left ( \frac{\hat p_\phi}{I}-\frac{\hbar \omega_0 r}{I cn_0} (\hat a_+^\dag \hat a_+ -\hat a_-^\dag \hat a_-)  \right )t \right ) {\hat a}_\pm(0).\nonumber
\end{align}

\subsection{Origin of SQL and possibility of back action evading measurement in the rotating system}\label{origin}

There are several possible problems that can be considered with respect of the system to explain the SQL introduced in Eqs.~(\ref{phisql}) and (\ref{omsql}). In what follows we will consider two of them. One is related to the measurement of initial angular velocity of rotating gyroscope, while the other is related to the measurement of the velocity change occurring during the measurement procedure. In what follows we argue that in the first case the accuracy of the measurement can be infinite, in accordance with Eq.~(\ref{omsql}), while in the second case it is limited by the standard quantum limit of phase detection, Eq.~(\ref{phisql}). Interestingly, selecting a proper measurement procedure one can remove the quantum back action in the second case.

Let us consider an empty open cavity rotating with angular velocity $\Omega$. The cavity is adiabatically interrogated with two, cw and ccw, optical pulses. Since operator $\hat p_\phi$ is conserved during the interaction, the angular momentum (and angular velocity) will be the same at the end of the measurement as the angular velocity at the beginning of the measurement. In other words $\hat p_\phi$ is an integral of motion, hence, it fulfils condition to  be  QND variable \cite{braginskykhalili_book}. Accuracy of our measurement procedure is limited by SQL, Eq.~(\ref{omsql}), though.

The information about the initial angular momentum, $\hat p_\phi (t=0)$, is contained in the phases $\varphi_\pm$ of the pulses exited the cavity \eqref{apm2}:
\begin{align}
 \varphi_\pm &= \mp\frac{r \Omega \omega_0 \tau}{cn_0}
\end{align}
Here $\tau$ is time duration of each pulse. The phase shift $(\varphi_--\varphi_+)/2$ can be measured with certain accuracy. For the simplest case when optical pulses are in the coherent state  with phase uncertainties $\Delta\varphi_\pm\simeq 1/(2\sqrt{n_\pm})$ and equal mean photon numbers $n_\pm=n$ we estimate the measurement error to be
\begin{align}
\label{Omegam}
 \Delta\Omega_\text{meas}& = \frac{cn_0}{2\omega_0r\tau}\sqrt{\Delta\varphi_+^2 +\Delta \varphi_-^2}
   \simeq \frac{cn_0}{2\omega_0r\tau\sqrt{2n}}
\end{align}

On the other hand, during measurement we get information on {\em perturbed} angular velocity, as it follows from
\eqref{dotphi},
\begin{align}
   \label{Omegaba}
 \Delta \Omega_\text{ba} &= \frac{\hbar \omega_0 r}{I cn_0} \big(\Delta n_+ + \Delta n_-\big)
   \simeq \frac{\hbar \omega_0 r}{I cn_0}\cdot 2\sqrt n
\end{align}
Here we again assume that optical pulses are in the coherent state with uncertainties of phonon numbers $\Delta n_\pm\simeq \sqrt n$.

The minimal error of the measurement can be derived from (\ref{Omegam}, \ref{Omegaba})
\begin{align}
   \label{DeltaO}
 \Delta\Omega_\text{min} &= \sqrt\frac{ 2\hslash}{I\tau},\quad
   \Delta p_\phi  = \sqrt\frac{ 2\hslash}{\tau}
\end{align}
It differs from SQL \eqref{omsql} by a numeric multiplier only.

Obviously, even within the SQL boundaries we can improve the accuracy of the measurement by increasing the measurement time $\tau$ or by utilizing the consequence of repeating measuring pulses. The latest technique simply allow involving multiple optical pulses to increase the measurement accuracy. Let us assume that during the procedure the measurement accuracy of the angular momentum is $\Delta_{p} $. The measurement can be repeated $N$ times. Since each measurement does not disturb the initial angular momentum and the errors of the measurements are not correlated, the accuracy of the set of measurements becomes $\Delta_{p} /\sqrt{N}$. The overall accuracy of the set of measurements increases with $N$ increase.

While the back action in operator $\hat p_\phi$ is removed after the measurement, the perturbation of the phase cannot be removed. This perturbation occurs in accordance with with \eqref{hatphi} (compare with \eqref{Omegaba}) and leads to
\begin{align}
\label{phip}
 \Delta \phi_\text{ba} &\simeq \frac{\hbar \omega_0 r\tau }{I cn_0}\cdot\sqrt{\Delta n_+^2 +\Delta n_-^2}
    \simeq \frac{\hbar \omega_0 r \tau \sqrt{2n}}{I cn_0}
\end{align}

It is possible to derive the standard uncertainty relationship for the measurements using (\ref{Omegam}, \ref{phip}):
\begin{align}
 \label{uncertainty1}
 I\Delta\Omega_\text{meas} \cdot \Delta \phi_\text{ba} \simeq \frac{\hslash}{2}.
\end{align}
If we use the phase of the output light waves to detect a change of the angular velocity that happens due to action of an external torque, the measurement sensitivity will be limited by the SQL related to the phase, not the angular velocity. We consider this case in the next section.

Depending on the measurement procedure the SQL can be lifted. The SQL appears in the QND measurement of the angular velocity because the back action \eqref{Omegaba} is erased {\em after} the measurement took place and initial $\hat p_\phi$ is not disturbed (it is QND variable) at the end. However, {\em during} the measurement it restricts the accuracy. In order to realize QND measurement along with back action evading (BAE) measurement  we have to use not the semi-classical coherent state, but a specifically prepared quantum state, preparation of which will be discussed somewhere else. We also can surpass SQL applying procedure of variational measurement \cite{96JetpVyMa, 96JetpVyMab, 01PRDKiLeMaThVy}, see also discussion and formula \eqref{xi} in what follows.

There is another simple thought experiment showing a possibility of BAE measurement. Let us consider a lossless system in which the rotating body is interrogated with a pair of optical pulses. The pulses enter and leave the optical cavity adiabatically. Let us imagine now, that instead of processing the pulses after the interaction one place an ideal mirror that reflects the pulses back to the cavity. Now the cw pulse becomes ccw and vice versa. When the pulses exit the cavity, they will have their state identical with their initial state. Both the information about cavity rotation and photon number in the pulses will be {\em completely erased} from the phases of the pulses.

In this configuration we will not be able to evaluate the initial rotation of the cavity. On the other hand, if the angular velocity of the cavity changes in the time interval between the pulses exit and re-entrance of the cavity, the resultant phase shift acquired by the pulses after the second interaction will be proportional to the angular velocity change. It will not be contaminated by the information about photon number in the pulse. In this way one should be able to measure the change of the rotation with unlimited accuracy that can be increased with increase of the photon number of the pulses in accordance with \eqref{Omegam}.

\subsection{Continuous measurement of torque}

In previous subsection we briefly mentioned a continuous measurement of a change of angular velocity of the system performed by detection of the phase of the light interacting with the rotating cavity. Let us consider this problem in more detail and study the accuracy of detection of classical torque acting on the ring cavity in the opto-mechanical system and consider an open lossless optical configuration by introducing coupling rate $\gamma$ and associated Langevin terms into the optic subsystem-related equations. We consider continuous measurement of torque acting on the cavity and assume that i) the probe cw and ccw light is resonantly tuned, and ii) the mechanical system stays without dissipation. The equations of motion describing the behavior of the {\em open } system become
\begin{subequations}
 \label{init2}
 \begin{align}
  \dot {\hat \phi} & = \frac{\hat p_\phi}{I}-\frac{\hbar \omega_0 r}{I cn_0}
    \big(\hat a_+^\dag \hat a_+ -\hat a_-^\dag \hat a_-\big) +\int \frac{T_s}{I}\, dt, \\
  \dot {\hat a}_\pm & + (i\omega_0+\gamma) {\hat a}_\pm  =\pm i \frac{r \dot {\hat \phi}}{cn_0}  \hat a_\pm
  + \sqrt\frac{2\gamma}{\tau} \hat b_\pm ,\\
  \hat d_\pm &= - \hat b_\pm +\sqrt{2\gamma\tau}\, {\hat a}_\pm.
\end{align}
\end{subequations}
Here $T_s$ is a time dependent signal torque, $\hat b_\pm$ are the amplitudes of the cw and ccw pump fields including fluctuation (Langevin) terms, $\tau =2\pi r/c$ is round trip time for the ring cavity, $\hat d_\pm$ are output amplitudes of the cw and ccw waves to be analyzed.

We remove the fast oscillating terms and present the optical field operators as sums of classical and quantum terms
\begin{align}
\hat a_\pm e^{i\omega_0 t} =A_\pm + a_\pm, \\
\hat b_\pm e^{i\omega_0 t} =B_\pm + b_\pm, \\
\hat d_\pm e^{i\omega_0 t} =D_\pm + d_\pm,
\end{align}
where $A_\pm$ are the field amplitudes inside the cavity, $B_\pm$ and $D_\pm$ are the input and output field mean amplitudes. For the sake of simplicity we assume that the mean amplitudes are real numbers  and write
\begin{align}
 A_\pm &= \sqrt\frac{2}{\gamma\tau} \cdot B_\pm,\\
 D_\pm & = B_\pm -\sqrt{2\gamma\tau} \,  A_\pm = - B_\pm\,,
\end{align}

We derive a set of equations in linear approximation for fluctuation amplitudes
\begin{subequations}
\label{Eqq3}
 \begin{align}
  \label{Eqq3+b}
  \dot a_+ & + \gamma a_+ = i\omega_0\cdot \frac{ r \dot \phi}{n_0 c}\, A_+ +\sqrt\frac{2\gamma}{\tau} b_+,\\
  \label{Eqq3-b}
  \dot a_- & + \gamma a_- = -i\omega_0\cdot \frac{ r \dot \phi}{n_0 c}\, A_- + \sqrt\frac{2\gamma}{\tau} b_-,\\
  \label{Eq3phib}
  I\, \ddot \phi & + \frac{\sqrt 2 \hslash\omega_0 r}{n_0 c} \partial_t
	 \left(A_+ a_\text{a+} - A_- a_\text{a-}\right)=T_s.\\
  & d_\pm = b_\pm -\sqrt{2\gamma\tau} \,  a_\pm .
  \end{align}
\end{subequations}
Amplitude noise components $a_\text{a+}$ and $a_\text{a-}$ are defined by Eq.~(\ref{apm}).

The equations for the fluctuations can be solved in frequency domain using Fourier transform if we neglect by the initial conditions of the opto-mechanical system:
\begin{subequations}
\label{EqqF}
 \begin{align}
  \label{EqqF1}
  a_\pm(t) &= \int_{-\omega_0}^\infty \alpha_\pm(\omega)\, e^{-i\omega t }\, d\omega,\\
  b_\pm(t) &= \int_{-\omega_0}^\infty \beta_\pm(\omega)\, e^{-i\omega t}\, d\omega,\\
  d_\pm(t) &= \int_{-\omega_0}^\infty \delta_\pm(\omega)\, e^{-i\omega t}\, d\omega,\\
  \label{EqqF2}
  \Omega(t)  &= \int_{-\infty}^\infty \Omega(\omega)\, e^{-i\omega t}\, d\omega,
  \end{align}
  \end{subequations}
While omitting the initial conditions for the optical modes is substantiated for the optical amplitudes in the case of a continuous measurement, it is not straightforward for the case of the mechanical degree of freedom. We use the motion that the initial conditions can be removed at the stage of the processing the classical signal taken during the measurements \cite{braginsky03prd}.

For the operators and their Fourier amplitudes the usual commutation relations are valid, for example
\begin{align}
\big[ a_\pm(t), a_\pm^\dag (t') \big]&=\delta(t-t'),\\
 \big[\alpha(\omega), \alpha^\dag(\omega')\big] &= 2\pi \delta(\omega -\omega')
\end{align}
Similar relationships are valid for the other operators.

It is convenient to introduce amplitude and phase quadratures for the fields
 \begin{subequations}
  \label{EqqQ}
  \begin{align}
  a_\text{a$\pm$}(t) &\equiv \frac{a_\pm(t) + a_\pm^\dag(t)}{\sqrt 2}, \label{apm}\\
		&a_\text{ph$\pm$} \equiv \frac{a_\pm(t) - a_\pm^\dag(t) }{i\sqrt 2}\, ,\\
  \alpha_\text{a$\pm$}(\omega) &\equiv \frac{\alpha_\pm(\omega) + \alpha_\pm^\dag(-\omega)}{\sqrt 2},\\
		& \alpha_\text{ph$\pm$} \equiv \frac{\alpha_\pm(\omega) - \alpha_\pm^\dag(-\omega) }{i\sqrt 2}
  \end{align}
 \end{subequations}
and rewrite equations \eqref{Eqq3} for the quadratures in frequency domain
\begin{subequations}
\label{Eqq4}
 \begin{align}
  \label{Eqq4+b}
  \alpha_\text{a$\pm$} &= \sqrt\frac{2\gamma}{\tau} \frac{\beta_\text{a$\pm$}}{\gamma-i\omega},\\
  \alpha_\text{ph$\pm$} &= \pm \frac{\sqrt 2 \omega_0 r\Omega}{n_0 c}\,\frac{ A_\pm}{\gamma-i\omega}
	 +\sqrt\frac{2\gamma}{\tau} \frac{\beta_\text{ph$\pm$}}{\gamma - i\omega},\\
  \label{Eq4phib}
  -i\omega &\left( I\, \Omega  +  \frac{\sqrt 2 \hslash\omega_0 r}{n_0 c}   \right. \times\\
  &\qquad \times
	 \left. \left(A_+\alpha_\text{a+} - A_- \alpha_\text{a-}\right)\right)=T_s,\nonumber
  \end{align}
\end{subequations}

For the output waves we obtain
\begin{subequations}
\label{Eqq5}
 \begin{align}
  \label{Eqq5apm}
  \delta_\text{a$\pm$} &= -\beta \beta_\text{a$\pm$},\quad \beta\equiv \frac{\gamma + i\omega}{\gamma -i\omega}\\
  \label{Eqq5d+}
  \delta_\text{ph$\pm$} &= \mp \frac{2\sqrt {\gamma\tau} \omega_0 r\Omega}{n_0 c}\,\frac{ A_\pm}{\gamma - i\omega}
	 - \beta\,\beta_\text{ph$\pm$},\\
  \label{Eq5phib}
   \Omega & = \frac{i T_s}{\omega\, I} - \frac{\sqrt 2 \hslash\omega_0 r}{ I n_0 c}
	  \left(A_+\alpha_\text{a+} - A_- \alpha_\text{a-} \right).
  \end{align}
\end{subequations}

After substitution \eqref{Eq5phib} into (\ref{Eqq5d+}) we obtain
 \begin{align}
 \nonumber
  \delta_\text{ph$\pm$} &= \pm \frac{2\sqrt {2 \gamma\tau} \hbar (\omega_0 r)^2}{(n_0 c)^2I}\,\frac{ A_\pm (A_+\alpha_\text{a+} - A_- \alpha_\text{a-})}{\gamma - i\omega} \\ \label{Eqq5zd+}
  &\mp  \frac{2\sqrt {\gamma \tau} \omega_0 r A_\pm}{ n_0 c(\gamma - i\omega)} \frac{ iT_s}{\omega\, I}
	 - \beta\,\beta_\text{ph$\pm$}.
  \end{align}
These equations can be rewritten with respect of the input fields

\begin{subequations}
\label{Eqq6}
 \begin{align}
  \delta_\text{a$\pm$} &= -\beta \beta_\text{a$\pm$}, \\
  \nonumber
  \delta_\text{ph$\pm$} &= \pm \frac{8 \hbar (\omega_0 r)^2}{(n_0 c)^2I\tau}\,\frac{ B_\pm (B_+\beta_\text{a+} - B_- \beta_\text{a-})}{(\gamma - i\omega)^2} \\ \label{Eqq6d+}
  &\mp  \frac{2\sqrt {2} \omega_0 r B_\pm}{ n_0 c(\gamma - i\omega)} \frac{ iT_s}{\omega\, I}
	 - \beta\,\beta_\text{ph$\pm$}.
  \end{align}
\end{subequations}
As expected, we see that the phase quadrature is contaminated with the optical power-dependent fluctuation term that comes along with the signal.

Let us consider measurement procedure of $T_s$ in which the sum and difference of outputs is detected
\begin{subequations}
\label{Eqq7}
 \begin{align}
   \tilde d_+ &= \frac{d_+ + d_-}{\sqrt 2},\quad \tilde d_- = \frac{d_+ - d_-}{\sqrt 2},\\
   \tilde b_+ &= \frac{b_+ + b_-}{\sqrt 2},\quad \tilde b_- = \frac{b_+ - b_-}{\sqrt 2},
 \end{align}
 \end{subequations}
For the sake of simplicity we also assume that the measurement is completely balanced $B_\pm=B$.
In frequency domain we obtain:
\begin{subequations}
\label{Eqq8}
  \begin{align}
	 \tilde \delta_\text{a$\pm$} &= -\beta \tilde \beta_\text{a$\pm$}, \quad
	 \tilde \delta_\text{ph+} = -\beta \tilde \beta_\text{ph+}\, ,\\ \nonumber
	 \tilde \delta_\text{ph--} &=  \beta \left(
	 \frac{16\hslash (\omega_0 r)^2\,  B^2  }{
		  I (n_0 c)^2\tau (\gamma^2 + \omega^2)}\,\tilde \beta_\text{a--}
	 - \tilde \beta_\text{ph--}\right) -\\ \nonumber
	 &\qquad -\frac{4 \omega_0 r B}{ n_0 c(\gamma - i\omega)}\cdot\frac{ iT_s}{\omega\, I}=\\
	 \label{deltaph-}
	 &=  \beta \left(\mathcal K \tilde \beta_\text{a--}
	 	 - \tilde \beta_\text{ph--}\right) - i \sqrt{2\beta \mathcal K}\frac{ T_s}{T_{SQL}},\\
	  &\mathcal K\equiv \frac{16\hslash (\omega_0 r)^2\, B^2 }{
		  I (n_0 c)^2\tau (\gamma^2 + \omega^2)}\ ,\\
	& T_{SQL}(\omega)\equiv \sqrt{2\hslash \, I\omega^2 }
  \end{align}
\end{subequations}
Here quadrature $\tilde \beta_\text{ph--}$ is responsible for measurement error while quadrature $\tilde \beta_\text{a--}$ is responsible for back action in the measurements.
In accordance with Eq.~(\ref{deltaph-}) the torque can be measured if
\begin{equation}
\frac{ T_s}{T_{SQL}(\omega)} > \sqrt{\frac{\mathcal K}{2}+\frac{1}{2\mathcal K} }.
\end{equation}
The minimum detectable torque is $T_s=T_{SQL}$ when $\mathcal K=1$ if one detects quadrature $\tilde \delta_\text{ph-}$.  We see that this straightforward measurement of the torque has limited maximum sensitivity even though an optimal strategy for the angular velocity measurement has unlimited sensitivity.

In order to surpass the SQL we can apply the procedure of a variational measurement \cite{96JetpVyMa, 96JetpVyMab, 01PRDKiLeMaThVy} of quadrature $\xi=\tilde \delta_\text{a--}\cos\theta +\tilde \delta_\text{ph--}\sin \theta$ defined as
\begin{align}
 \label{xi}
 \xi &= \beta \left(\big[\mathcal K \sin\theta + \cos\theta\big]\tilde \beta_\text{a-}
	 	 - \sin\theta\tilde \beta_\text{ph--}\right)-\\
	 	 &\qquad - i\sin\theta \sqrt{2\beta \mathcal K}\frac{ T_s}{T_{SQL}},\nonumber
\end{align}
where homodyne angle $\theta$ is selected so that $\cot \theta= -1/\mathcal K$. Such a selection allows for compensation of the back action. Due to frequency dependence of $\mathcal K$ this compensation is possible in a limited frequency band only. Similar method is used in conventional quantum speed meter to overcome the SQL restriction \cite{90PHABrKh, 00PRDBrGoKhTh}.

\section{Discussion}

In this paper we analyze fundamental restrictions of quantum sensitivity of gyroscope
and show, the gyroscope should be extremely small in size to allow these quantum effects to emerge.
Microring optical cavities can have high finesse and can be monolithically integrated. It gives us grounds to expect that the quantum effects can be observed with this type of devices. On the other hand, the demonstrated measurement sensitivity of the existing micro-gyroscopes  \cite{suzuki00jlt,Sanders06ofs,Mao11oe,Ma11jlt,Srinivasan14oe,Ma15oe,Wang15ol,Ma15ol,venediktov16qe} is not very high. It was shown that an on-chip gyroscope having resolution on the order of $10$~deg/hr is feasible if an InP-based ring cavity with Q factor of approximately $6\times10^5$ and footprint of 10~mm$^2$ is utilized  \cite{ciminelli16pj}. An active on-chip gyroscope having similar resolution has also been demonstrated \cite{li17o}. A resonant gyroscope \cite{liang16iss,liang17o} utilizing ultra-high-Q crystalline whispering gallery mode resonators and having $2$~deg/hr resolution was also demonstrated. The sensitivity should be improved significantly to allow quantum effects described in this paper become visible.

\begin{figure}
 \includegraphics[width=0.4\textwidth]{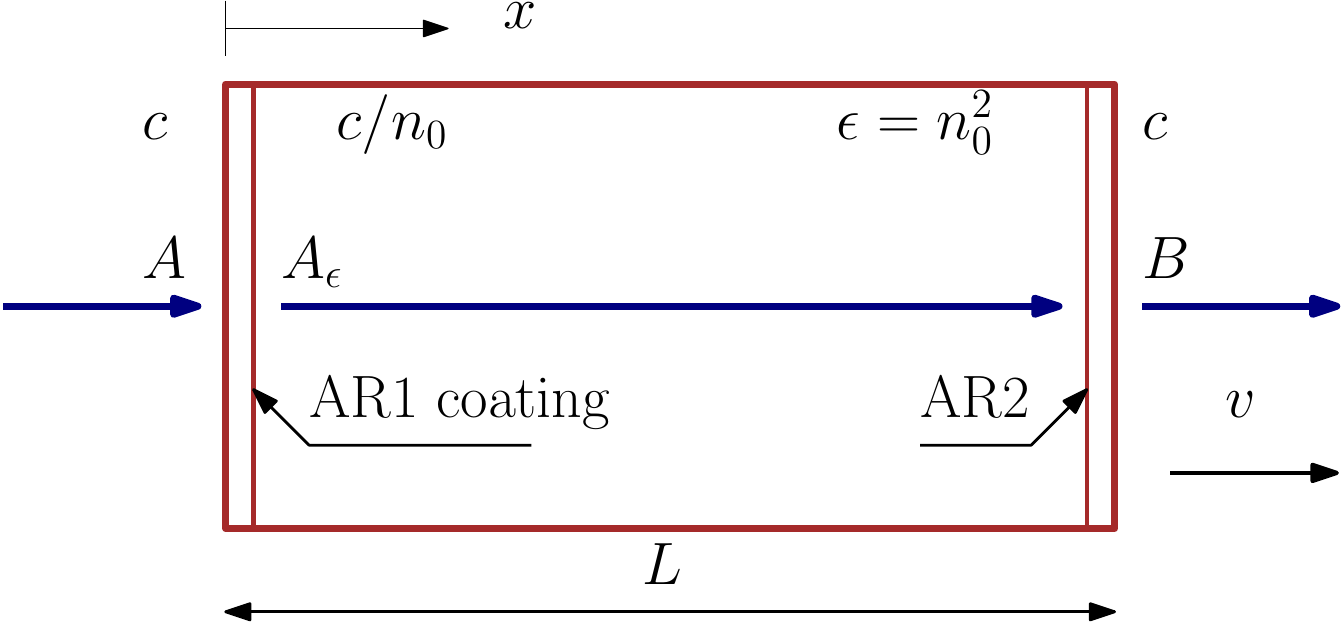}
 \caption{A toy model of speed meter. The velocity $v$ of a dielectric (refractive index is $n_0$) probe mass with anti-reflecting  coatings (AR1, AR2) is measured by  light traveling through it via detection of phase shift of the light proportional to velocity $v$.}\label{arSM}
\end{figure}

The formalism developed in this paper also can be useful for measurements performed with straight moving bodies. The measurement of the angular velocity is similar to the measurement of velocity of a body. The analogy becomes obvious if one considers, for instance, a rotating gyroscope as a pointer mass that can move on a round string. Let us show here the analogy introducing a toy semi-classical model of the speed meter shown in Fig.~(\ref{arSM}).

The light passes through a dielectric test mass without reflections (its surfaces are covered by anti-reflecting coatings), and its phase $\varphi$ shift depends on velocity $v$ of the test mass (see notations Fig.~\ref{arSM})
\begin{align}
 \phi &= \frac{\omega_0}{c}\, v\tau_0(n_0-1), \quad \tau_0 = \frac{n_0L}{c}
\end{align}
Here $\omega_0$ is the optical frequency, $n_0$ is the refractive index of the material.
Let us assume that the light is in the coherent state. The error of the velocity measurement is
\begin{align}
\label{vm}
 \delta \phi &\simeq \frac{1}{2\sqrt{n}},\quad \Delta v_\text{meas} \simeq \frac{1}{2\sqrt{n}}\cdot \frac{c}{\omega_0\tau(n_0-1)}
\end{align}
Here $n$ is number of ``used'' (passed through) light quanta and $\tau$ is the duration of the measurement.

The momentum of the optical quanta inside and outside of the test mass are equal to
\begin{align}
 p_\text{out} &= \frac{\hslash \omega_0}{c},\quad
    p_\text{inside} = \frac{n_0 \hslash \omega_0}{c}\,,
\end{align}
respectively.  Hence, while photon travelling inside, the probe mass receives additional momentum $  p_\text{out} -  p_\text{inside}$, which transforms into position shift  of the probe mass
\begin{align}
 x = \frac{ p_\text{out} -  p_\text{inside}}{m}\cdot \tau
\end{align}
The uncertainty of quanta number $\sqrt n$ produces the back action noise of the velocity of the probe mass ($m$)during the measurement
\begin{align}
\label{vba}
 \Delta v_\text{ba} &\simeq \frac{\sqrt{n}\, \hslash \omega_0 (n_0-1)}{c m}
\end{align}
This perturbation is erased after the measurement because of the photon number and energy conservation laws.

Combining (\ref{vm} and \ref{vba}) one obtains minimal error of the measurement
\begin{align}
 \Delta v_\text{min} &=\sqrt\frac{\hslash}{m\tau}
\end{align}
which coincides with the SQL for the velocity. We can surpass SQL applying procedure of variational measurement \cite{96JetpVyMa, 96JetpVyMab, 01PRDKiLeMaThVy}.

The uncertainty of quanta number $\sqrt n$ produces the uncertainty of position (back action):
\begin{align}
\label{xba}
 \Delta x_\text{ba} &\simeq \frac{\sqrt{N}\, \hslash \omega_0 (n_0-1)}{c m}\cdot \tau
\end{align}
Leading to uncertainty relation
\begin{align}
\label{uncertainty2}
   m \Delta v_\text{meas}\cdot \Delta x_\text{ba}\simeq \frac{\hslash}{2}
\end{align}
Formulas (\ref{vm}, \ref{xba}, \ref{uncertainty2}) here directly relates to formulas (\ref{Omegam}, \ref{phip}, \ref{uncertainty1}) for quantum gyroscope correspondingly.

The back action evading QND measurement strategy introduced at the end of SubSec.~\ref{origin} is also valid for the speed meter. Let us consider light pulse  passing through the test mass, then reflecting from a perfect mirror and passing the test mass in the opposite direction. During the second propagation the light-induced ponderomotive force  has the opposite direction and is completely erased. The accuracy of test mass velocity is defined only by \eqref{vm} and can be decreased just by increase of the photon number $n$ of the optical pulse.

\section{Conclusion}

We have shown that the sensitivity of a generalized gyroscope is restricted by the standard quantum limit, in a way similar to the free mass coordinate measurement sensitivity limitations. Quantum theory indicates that the detection of the rotation rate and rotation phase are fundamentally different because the rotation phase and the canonical angular momentum operators do not commute. As the result, the observables cannot be measured simultaneously with high accuracy. Using an example of a resonant optical gyroscope we have found the ultimate limit of the sensitivity of the device and discussed requirements for achievement of the sensitivity in an experiment. A back action evading measurement technique allowing surpassing the standard quantum limit is proposed. We also analyse quantum restrictions of sensitivity of measurements of the classical torque applied to the gyroscope.

\acknowledgments

S.V. acknowledges partial support from  Russian Science Foundation (Grant No. 17-12-01095) and from the TAPIR gift funds for MSU support.

\end{document}